\documentclass[twocolumn]{article}

\usepackage{amssymb}
\usepackage{amsfonts}

\newtheorem{theorem}{Theorem}

\newtheorem{example}[theorem]{Example}

\begin{document}

\title{
Reverse estimation theory, Complementarity between SLD and RLD,
and monotone distances
}
\author{
Keiji Matsumoto$^{\ast\dagger}$
}

\maketitle

\footnotetext[1]{National Institute of Informatics\\
2-1-2, Hitotsubashi, Chiyodaku, \\
Tokyo 101-8430\\
TEL:+81-3-4212-2560 FAX: +81-3-3556-1916\\
}
\footnotetext[2]{
Quantum Computation and Information Project, JST\\
5-28-3 Hongo, Bunkyo-ku, Tokyo 113-0033\\
TEL: +81-3818-3314 
}

\begin{abstract}
Many problems in quantum information theory 
can be vied as interconversion between resources.
In this talk, we apply this view point to 
state estimation theory, motivated by the following observations.

First, 
a monotone metric takes value between SLD and RLD Fisher metric.
This is quite analogous to the fact that entanglement measures
are sandwiched by distillable entanglement and entanglement cost.
Second, SLD add RLD are mutually complement
via purification of density matrices, 
but its operational meaning
was not clear.

To find a link between these observations, we define
reverse estimation problem, or simulation of 
quantum state family by probability distribution family,
proving that RLD Fisher metric is a solution to 
local reverse estimation problem of quantum state family
with 1-dim parameter. 
This result gives new proofs of some known facts 
and proves one new fact about monotone distances.

We also investigate information geometry of RLD,
and reverse estimation theory of a multi-dimensional parameter family.
\end{abstract}

\section{Introduction}

Many problems in quantum/classical information theory can be viewed as
interconversion between a given resource and a 'standard' resource, and such
viewpoint had turned out to be very fruitful. This manuscript will exploit\
this scenario in asymptotic theory of quantum estimation theory (with some
comments on classical estimation theory).

Resource conversion scenario was first explored in axiomatic theory of
entanglement measures. Entanglement is a kind of quantum non-locality, which
cannot be explained by classical mechanical theory. Its effect is typically
visible in so called maximally entangled states, which we regards as
standard resources. The optimal asymptotic conversion ratio from maximally
entangled state to a given state is called \textit{entanglement cost}, while
the optimal ratio for inverse conversion is called \textit{distillable
entanglement}. It is shown that all quantities which satisfies a set of
reasonable axioms takes value between these two quantities.

It had been known that a monotone metric in quantum state space takes value
between RLD and SLD Fisher metric. It had been also pointed out that these
metrics are mutually 'complement', in the following sense: A mixed quantum
state can be regarded as a reduced state of pure state in larger system. In
this view, SLD Fisher metric of a quantum state space equals RLD Fisher
metric in the space of quantum states in ancilla system.

In the manuscript, we link these two facts via resource conversion scenario,
giving new proof of the former, and monotonicity of RLD and SLD Fisher
metrics. We also prove similar statement for quantum version of relative
entropy, which, to author's knowledge, is new. In the discussion, estimation
corresponds to distillation of standard resource.

In above discussion, we need 'reverse estimation', which we formulate as 
\textit{reverse estimation }of quantum state families: Given a family of
quantum state, its reverse estimation is a CQ map and a family of
probability distributions such that the output of the CQ map equals the
quantum state family (In fact, we mainly consider local version of this, to
make a complement to local estimation theory, which is equivalent to
asymptotic estimation theory).

Next, we study local reverse estimation itself more in detail. Especially,
we point out that, in general, local 'reverse estimation' is mathematically
equivalent to local estimation with fixed set of observables.
Straightforward calculation shows that optimal reverse estimation
corresponds to P-representation. For the Q-representation corresponds to
optimal estimation, gap between reverse estimation and estimation origins
from uncertainty contained in coherent states.

\section{SLD and RLD, Fisher information}

In the manuscript, we restrict ourselves to finite (namely, $d$- )
dimensional Hilbert space ${\cal H}$, unless otherwise mentioned. The
totality of density matrices is denoted by ${\cal S}\left( {\cal H}%
\right) $, and the totality rank $r$ elements is denoted by ${\cal S}%
_{r}\left( {\cal H}\right) $. In the manuscript, $r=d$, unless otherwise
mentioned. Unless otherwise mentioned, a parameterized family of quantum
states, often denoted by ${\cal M}=\left\{ \rho _{\theta };\theta \in
\Theta \subset \mathbb{R}^{m}\right\} $, 
is assumed to be differentiable up to an arbitrary order.

Define a \textit{symmetric logarithmic derivative} $L_{\theta ,i}^{S}$and a 
\textit{right logarithmic derivative} $L_{\theta ,i}^{R}$ as a solution to
the matrix equation,%
\[
\partial _{i}\rho _{\theta }=\frac{1}{2}(L_{\theta ,i}^{S}\rho _{\theta
}+\rho _{\theta }L_{\theta ,i}^{S})=L_{\theta ,i}^{R}\rho _{\theta }, 
\]%
where $\partial _{i}:=\frac{\partial }{\partial \theta ^{i}}$. If $\rho
_{\theta }$ is strictly positive, $L_{\theta ,i}^{S}$ and $L_{\theta ,i}^{R}$
are uniquely defined in this way. If $\rho _{\theta }$ has zero eigenvalues, 
$L_{\theta ,i}^{S}$ still can be defined, but not uniquely. $L_{\theta
,i}^{R}$ exists (and if exists, unique) if and only if $\partial _{i}\rho
_{\theta }$ has non-zero eigenvalues only in the support of $\rho _{\theta }$%
. Observe they are quantum equivalences of a classical logarithmic
derivative, $\partial _{i}\log p_{\theta }(x)$.

An SLD Fisher information matrix $J_{\theta }^{S}$ and RLD Fisher
information matrix $J_{\theta }^{R}$ are defined as%
\begin{eqnarray*}
J_{\theta ,i,j}^{S} &=&\Re {\rm Tr}\rho _{\theta }L_{\theta
,i}^{S}L_{\theta ,j}^{S}, \\
J_{\theta ,i,j}^{R} &=&{\rm Tr}\rho _{\theta }L_{\theta ,j}^{R\dagger
}L_{\theta ,i}^{R}.
\end{eqnarray*}%
They are quantum analog of a classical Fisher information matrix,

\[
J_{\theta ,i,j}:=\sum_{x}p_{\theta }(x)\partial _{i}\log p_{\theta
}(x)\partial _{j}\log p_{\theta }(x). 
\]%
For they are positive definite, they can be regarded as metric tensors
introduced to the tangent space $T_{\theta }\left( {\cal M}\right) $, and
the corresponding metrics are called SLD Fisher metric and RLD Fisher
metric,respectively.

\section{Duality between SLD and RLD, Reverse SLD}

Denote by ${\cal W}$ the totality of matrices with finite numbers of rows
and columns. In the manuscript, an element of ${\cal W}$ is considered as
an ordered set of unnormalized state vectors which forms a convex
decomposition of a mixed state, with their magnitudes' corresponding to
weights. Equivalently, an element of ${\cal W}$ can be interpreted as a
representation of bipartite pure state, whose reduced density matrix to one
of the parties equals a given density matrix. The totality of $d\times
d^{\prime }$ elements of ${\cal W}$ is denoted by ${\cal W}_{d^{\prime
}}$.

We consider a map $\pi $ form ${\cal W}$ to ${\cal S}\left( {\cal H}%
\right) $,%
\[
\pi :W\rightarrow WW^{\dagger }. 
\]%
An interpretation of this map is as follows. Let 
\[
W=[\sqrt{p_{1}}\left\vert \phi _{1}\right\rangle ,\cdots ,\sqrt{p_{d^{\prime
}}}\left\vert \phi _{d^{\prime }}\right\rangle ], 
\]%
then, 
\[
\pi (W)=\sum_{i=1}^{d^{\prime }}p_{i}\left\vert \phi _{i}\right\rangle
\left\langle \phi _{i}\right\vert . 
\]%
Another interpretation would be given by taking correspondence, 
\[
\left\vert \Phi _{W}\right\rangle =\sum_{i,j}w_{i,j}\left\vert
e_{i}\right\rangle \left\vert f_{j}\right\rangle , 
\]%
where $\{\left\vert e_{i}\right\rangle \}$ is an orthonormal basis in $%
{\cal H}$ , and $\{\left\vert f_{j}\right\rangle \}$ is an orthonormal
basis in a Hilbert space ${\cal H}^{\prime }$ for purification. Then, 
\[
\pi (W)={\rm Tr}_{{\cal H}^{\prime }}\left\vert \Phi _{W}\right\rangle
\left\langle \Phi _{W}\right\vert . 
\]

Its differential map is denoted by $\pi _{\ast }:T\left( {\cal W}\right)
\mapsto T\left( {\cal S}\left( {\cal H}\right) \right) $, where $%
T_{W}\left( {\cal W}\right) $ means a tangent space to ${\cal W}$ at $W
$, and $T\left( {\cal W}\right) $ is a tangent bundle, or the union of $%
T_{W}\left( {\cal W}\right) $, with $W$'s running all over ${\cal W}$.
An element of $T\left( {\cal W}\right) $ is naturally represented by an
element of ${\cal W}$ by considering a parameterized family of an
elements ${\cal W}$ and differentiating with respect to a parameter.
Denote such representation of an element $\widehat{X}$ of $T_{W}\left( 
{\cal W}\right) $ by ${\bf M}\widehat{X}$ , or more explicitely,
\[
{\bf M}\left. \frac{\partial }{\partial \zeta ^{i}}\right\vert_{W}
:=2\left. \frac{\partial W_{\zeta }}{\partial \zeta ^{i}}\right\vert_{W}.
\]%
In that representation, 
\[
\pi _{\ast }(X)=\frac{1}{2}\left\{ W\left( {\bf M}\widehat{X}\right)
^{\dagger }+\left( {\bf M}\widehat{X}\right) W^{\dagger }\right\} ,
\]%
which is easily understood recalling Leibnitz's rule of differentiation of a
product of two matrices.

Observe that these maps are not unique. First, the map $\pi $ satisfies,%
\[
\pi (WU)=\pi (W)
\]%
where $U$ is a matrix with $UU^{\dagger }={\rm I}$ (need not to be a
unitary). Sometimes, this transform is refered to as a gauge transform.
Correspondingly, the kernel of $\pi _{\ast }$, denoted by ${\cal K}%
_{W}\left( {\cal W}\right) $, is 
\[
{\cal K}_{W}\left( {\cal W}\right) {\cal =}\left\{ \widehat{X}\,;\,%
{\bf M}\widehat{X}=WA^{K},\,\exists A^{K}=-A^{K\dagger }\right\} .
\]

Denote an element of $T_{\rho }({\cal S}({\cal H}))$ by $X$, and
denote an SLD and RLD corresponding to $X$ by $L_{\rho ,X}^{S}$ and $L_{\rho
,X}^{R}$, respectively. We define two inverse maps of $\pi _{\ast }$, which
are denoted by $h_{W}^{S}$and $h_{W}^{R}$ (subscript $W$ is often dropped)
as,%
\begin{eqnarray*}
\,{\bf M}h_{W}^{S}(X) &=&L_{\rho ,X}^{S}W, \\
{\bf M}h_{W}^{R}(X) &=&L_{\rho ,X}^{R}W,
\end{eqnarray*}%
with $\rho =\pi \left( W\right) $. It is easy to verify $\pi _{\ast }\circ
h_{W}^{S}=\pi _{\ast }\circ h_{W}^{R}={\rm id}$. Consider subspaces $%
{\cal LS}_{W}\left( {\cal W}\right) $ and ${\cal LR}_{W}\left( 
{\cal W}\right) $ of $T_{W}\left( {\cal W}\right) $ which are defined
by,

\begin{eqnarray*}
{\cal LS}_{W}\left( {\cal W}\right)  =\left\{ \widehat{X}\,;\,%
\widehat{X}=h_{W}^{S}(X),\,\exists X\in T_{\pi (\rho )}\left( {\cal S}%
\left( {\cal H}\right) \right) \right\} , \\
{\cal LR}_{W}\left( {\cal W}\right)  =\left\{ \widehat{X}\,;\,%
\widehat{X}=h_{W}^{R}(X),\,\exists X\in T_{\pi (\rho )}\left( {\cal S}%
\left( {\cal H}\right) \right) \right\} .
\end{eqnarray*}%
It is easy to see%
\[
L_{\rho ,X}^{R}W=WA^{R},\,\exists A^{R}=A^{R\dagger }.
\]%
$A^{R}$ is said to be the \textit{reverse SLD at }$W$.

Define a map $\widetilde{\pi }$ from ${\cal W}$ to ${\cal S}({\cal H%
}^{\prime })$ such that,%
\[
\widetilde{\pi }(W)=W^{\dagger }W=\pi (W)={\rm Tr}_{{\cal H}%
}\left\vert \Phi _{W}\right\rangle \left\langle \Phi _{W}\right\vert .
\]%
Correspondingly, we can define 
$\widetilde{h}_{W}^{S}(X)$, $\widetilde{h}_{W}^{R}(X)$, \\
$\widetilde{{\cal LS}}_{W}\left( {\cal W}\right) $, and $%
\widetilde{{\cal LR}}_{W}\left( {\cal W}\right) $, for which 
\begin{eqnarray*}
\widetilde{{\cal LS}}_{W}\left( {\cal W}\right)  &\supset &{\cal LR}%
_{W}\left( {\cal W}\right) , \\
\widetilde{{\cal LR}}_{W}\left( {\cal W}\right)  &\supset &{\cal LS}%
_{W}\left( {\cal W}\right) 
\end{eqnarray*}%
holds. Especially, if $d^{\prime }=r$, 
the LHS and the RHS coinside with each other.
This means than \textit{RLD of the system corresponds to SLD of the
ancilla system}. Also, we have, 
\[
{\rm Tr}\rho L^{R\dagger }L^{R}\leq {\rm Tr}\tilde{\pi}(W)A^{R}A^{R}.
\]%
Especially, if $d^{\prime }=r$, the equality holds. These relations are
called \textit{duality} between SLD and RLD.

\section{Reverse estimation of quantum sate family and RLD}

The heart of quantum statistics is optimization of a measurement, i.e.,
choice of a measurement which converts a family of quantum states to the
most informative classical probability distribution family. Let us denote by 
$p_{\theta }^{M}$ the probability distribution of measurement results of $M$
applied to $\rho _{\theta }$, and denote by $J_{\theta }^{M}$ the classical
Fisher information matrix of the probability distribution family $%
\{p_{\theta }^{M}\}$. Then, it is known that, for a $1$-dim quantum state
family ${\cal M}$, 
\[
\max_{M:{\rm meas}.}J_{\theta }^{M}=J_{\theta }^{S}, 
\]%
or, $J_{\theta }^{S}$ is the maximal amount of classical Fisher information
extracted from the $1$-dim quantum state family $\{\rho _{\theta }\}$ at $%
\theta $. In other words, we consider a QC map which maximizes the output
Fisher information.

Now, we consider the reverse of above, i,e, emulation of the $1$-dim quantum
state family $\{\rho _{\theta }\}$ at $\theta _{0}$ up to the first order,
i.e., a pair $(\Phi ,\{p_{\theta }\})$ of the probability
distribution family $\{p_{\theta }\}$ such that with a QC channel $\Phi $,
such that,%
\begin{eqnarray}
\Phi \left( p_{\theta _{0}}\right) &=&\rho _{\theta _{0}},  \label{localsim}
\\
\left. \frac{{\rm d}\Phi \left( p_{\theta }\right) }{{\rm d}\theta }%
\right\vert _{\theta =\theta _{0}} &=&\left. \frac{{\rm d}\rho _{\theta }%
}{{\rm d}\theta }\right\vert _{\theta =\theta _{0}}.  \nonumber
\end{eqnarray}%
Our task is to optimize a pair $(\Phi ,\{p_{\theta }\})$, called local
reverse estimation at $\theta _{0}$, to minimize Fisher information $%
J_{\theta }$ of the input $\{p_{\theta }\}$.

A local reverse estimation of $\{\rho _{\theta }\}$ at $\theta _{0}$ is
constructed as follows. Define a system of state vectors $\left\vert \phi
_{1}\right\rangle ,\cdots ,\left\vert \phi _{d^{\prime }}\right\rangle $ ,
and a probability distribution $\{p(i)\}$ by the equations,%
\[
\rho _{\theta _{0}}=\sum_{x=1}^{d^{\prime }}p(x)\left\vert \phi
_{x}\right\rangle \left\langle \phi _{x}\right\vert , 
\]%
This corresponds to a QC map $\Phi $ which outputs $\left\vert \phi
_{x}\right\rangle $ according to the input probability probability
distribution $p(x)$. Define real numbers $\lambda _{1},\cdots ,\lambda
_{d^{\prime }}$ by%
\[
\left. \frac{{\rm d}\rho _{\theta }}{{\rm d}\theta }\right\vert
_{\theta =\theta _{0}}=\sum_{x=1}^{d^{\prime }}\lambda _{x}p(x)\left\vert
\phi _{x}\right\rangle \left\langle \phi _{x}\right\vert . 
\]%
and define $\{p_{\theta }\}$ by $p_{\theta }\left( x\right) :=p(x)+\lambda
_{x}p(x)(\theta -\theta _{0})$. Then, the pair $(\Phi ,\{p_{\theta }\})$ is
a local reverse estimation, and any local reverse estimation is given in
this way, essentially (i.e., modulo the difference of $o(|\theta -\theta
_{0}|)$).

Define also 
\begin{eqnarray*}
W &=&[\sqrt{p(1)}\left\vert \phi _{1}\right\rangle ,\cdots ,\sqrt{p(l)}%
\left\vert \phi _{d^{\prime }}\right\rangle ] \\
A &=&{\rm diag}(\lambda _{1,\cdots ,}\lambda _{d^{\prime }}).
\end{eqnarray*}%
Then , we have, $\left. \frac{{\rm d}\rho _{\theta }}{{\rm d}\theta }%
\right\vert _{\theta =\theta _{0}}=WAW^{\dagger }$, and 
\[
L_{\theta _{0},1}^{R}WP=WA, 
\]%
with $P$ being the projector onto the support of $\tilde{\pi}\left( W\right) 
$ . The logarithmic derivative of $\{p_{\theta }\}$ at $\theta =\theta _{0}$
is 
\[
\left. \frac{{\rm d\log }p_{\theta }(x)}{{\rm d}\theta }\right\vert
_{\theta =\theta _{0}}=\lambda _{x}, 
\]%
and its Fisher information is, 
\begin{eqnarray*}
J_{\theta _{0}} &=&\sum_{x=1}^{d^{\prime }}\left( \lambda _{x}\right)
^{2}p(x)={\rm Tr}WAAW^{\dagger } \\
&\geq &{\rm Tr}WAPAW^{\dagger }=J_{\theta _{0}}^{R}.
\end{eqnarray*}%
The equality holds if $P$ equals the identity, or $d^{\prime }=r$. Hence, to
simulate $\{\rho _{\theta }\}$ at the neighbor of $\theta _{0}$ up to the
first order, we need classical Fisher information by the amount of $%
J_{\theta _{0}}^{R}$.

\begin{theorem}
\[
\max J_{\theta _{0}}=J_{\theta _{0}}^{R}, 
\]%
where maximization is taken over all the local reverse estimations of $%
\{\rho _{\theta }\}$ at $\theta _{0}$.
\end{theorem}

\section{Monotone metric revisited}

It is known that SLD Fisher metric and RLD Fisher metric are monotone by
application of CPT map, and any monotone metric takes value between SLD and
RLD Fisher metric. In this section, we demonstrate operational meaning of
SLD and RLD implies these properties in trivial manner.

First, monotonicity of SLD is trivial because the optimization of
measurement applied to the family $\{\Lambda \rho _{\theta }\}$ is
equivalent to the optimization of measurement to $\{\rho _{\theta }\}$ over
all the restricted class of measurement of the form $M\circ \Lambda $.

The monotonicity of RLD Fisher metric is proven in the similar manner. Given
a local reverse estimation $(\Phi ,\{p_{\theta }\})$ of $\{\rho _{\theta }\}$
at $\theta _{0}$, $(\Lambda \circ \Phi ,\{p_{\theta }\})$ is a local reverse
estimation of the family $\{\Lambda \rho _{\theta }\}$ at $\theta _{0}$. We
may be able to improve this reverse estimation to reduce the amount of
classical Fisher information of the probability distribution family. Thus
the monotonicity of RLD Fisher metric is proved.

Also, we can prove that SLD Fisher metric is no larger than RLD by
considering composition of the optimal local reverse estimation followed by
the optimal measurement. This operation, being a CPT map, cannot increase
classical Fisher information. For the initial classical Fisher information
equals RLD Fisher information and the final one equals SLD Fisher
information, we have the inequality.

Assume that a metric is not increasing by a QC channel, and coincides with
classical Fisher information restricted to classical probability
distributions. Then, this metric should be no smaller than SLD Fisher
metric. Let us consider a $1$-dim family $\{\rho _{\theta }\}$. If one apply
an optimal QC map, classical Fisher information $J_{\theta }$ of the output
probability distribution family $\{p_{\theta }\}$ equals $J_{\rho _{\theta
}}^{S}$. Due to the latter assumption, $g_{p_{\theta }}=J_{\rho _{\theta
}}^{S}$. Therefore, the monotonicity  by a QC channel $g_{\rho _{\theta
}}\geq g_{p_{\theta }}=J_{\rho _{\theta }}^{S}$.

Similarly, assume that a metric is not increasing by a CQ channel and
coincides with classical Fisher information restricted to classical
probability distributions. Then, the metric should be no larger than RLD
Fisher metric.Consider an optimal local reverse estimation of the $1$-dim
family $\{\rho _{\theta }\}$ at $\theta $. Then, classical Fisher
information $J_{\theta }$ of the input probability distribution family $%
\{p_{\theta }\}$equals $J_{\theta }^{R}$. Due to the latter assumption, $%
g_{p_{\theta }}=J_{\rho _{\theta }}^{R}$. Therefore, the monotonicity  by a
CQ channel $g_{\rho _{\theta }}\leq g_{p_{\theta }}=J_{\rho _{\theta }}^{R}$.

Altogther, if a metric is monotone non-increasing by application of QC and
CQ maps, the metric takes value between SLD and RLD Fisher metric.

\begin{theorem}
Assume that a metric $g$ coincide with classical Fisher information in the
space of classical probability distributions. In addition, if $g$ is
monotone decreasing by a QC map, $g$ is larger than SLD Fisher metric. If $\
g$ is monotone decreasing by a CQ map, $g$ is smaller than RLD Fisher metric.
\end{theorem}

\section{Global reverse estimation}

Let us define a global reverse estimation of a quantum state family $\{\rho
_{\theta }\}$ is a pair $(\Phi ,\{p_{\theta }\})$ of  the
probability distribution family $\{p_{\theta }\}$ such that with a QC
channel $\Phi $, such that,%
\[
\Phi (p_{\theta })=\rho _{\theta },\forall \theta \in \Theta .
\]%
This is equivalent to%
\begin{eqnarray*}
\rho _{\theta } &=&\sum_{i}p_{\theta }(i)\left\vert \phi _{i}\right\rangle
\left\langle \phi _{i}\right\vert  \\
&=&W_{0}M_{\theta }W_{0}^{\dagger },
\end{eqnarray*}%
where $M_{\theta }=$ ${\rm diag}(p_{\theta }(x),\cdots ,p_{\theta }(x))$.
Let%
\begin{eqnarray*}
A_{\theta ,i}^{R} &=&{\rm diag}(\partial _{1}\log p_{\theta }(x),\cdots
,\partial _{r}\log p_{\theta }(x)), \\
W_{\theta } &=&W_{0}\sqrt{M_{\theta }}.
\end{eqnarray*}%
For 1-$\dim$ restriction of achives RLD Fisher infomation,
 we have to have,%
\begin{equation}
L_{\theta ,i}^{R}W_{\theta }=W_{\theta }A_{\theta ,i}^{R}.
\end{equation}

\begin{theorem}
If $\rho _{\theta }$ is a full-rank matrix for all $\theta \in \Theta $, the
following three are equivalent.

\begin{description}
\item[(i)] The state family $\left\{ \rho _{\theta }\right\} $ has a global
reverse estimation such that its $1$-$\dim $ restriction achieves RLD Fisher
information at all $\theta \in \Theta $.

\item[(ii)] $\rho _{\theta }=W_{\theta _{0}}M_{\theta }W_{\theta
_{0}}^{\dagger }$, where $M_{\theta }$ is a $r\times r$ Hermitian matrix,
and $[M_{\theta _{1}},M_{\theta _{2}}]=0$ for all $\theta _{1}$, $\theta _{2}
$.

\item[(iii)] $[L_{\theta ,i}^{R},\,L_{\theta ^{\prime },j}^{R}]=0$ for all $i
$, $j$, $\theta $, and $\theta ^{\prime }$.
\end{description}
\end{theorem}

\section{Two point reverse estimation}

Now, we turn to reverse estimation of two quantum states, $\rho $, $\sigma $%
, which is a pair $(\Phi ,\{p_{\lambda };\lambda =\rho ,\sigma \})$ of a CQ
map and a probability distribution family such that $\Phi (p_{\rho })=\rho $
and $\Phi (p_{\sigma })=\sigma $ . The problem discussed here is the
minimization of the divergence between the probability distributions between 
$p_{\rho }$ and $p_{\sigma }$. 

It is known that the divergence equals a integral of metric along a curve, $%
\{p_{t}^{(m)}=tp_{\rho }+(1-t)p_{\sigma }\}$, 
\[
D\left( p_{\rho }||p_{\sigma }\right) =\int_{0}^{1}\int_{0}^{t}J_{s}^{(m)}%
{\rm d}s{\rm d}t,
\]%
where $J_{t}^{(m)}$ is a Fisher information of the family $\{p_{t}^{(m)}\}$.
This quantity is upper-bounded by 
\begin{equation}
D^{R}(\rho ||\sigma ):=\int_{0}^{1}\int_{0}^{t}J_{s}^{R}{\rm d}s{\rm d}%
t,  \label{rld-divergence-int}
\end{equation}%
where $J_{t}^{R}$ the RLD Fisher information of the family of quantum states 
$\left\{ \Phi (p_{t}^{(m)})\right\} $. Observe that, for any reverse
estimation, we have 
\[
\Phi (p_{t}^{(m)})=\rho _{t}^{(m)}:=t\rho +(1-t)\sigma .
\]%
Hence, $D\left( p_{\rho }||p_{\sigma }\right) $ is maximized if $J_{t}^{(m)}$
is maximized at each $t$. If $J_{t}^{R}=J_{t}^{(m)}$ for all $t$ ($0\leq
t\leq 1$), i.e., the reverse estimation is an optimal local reverse
estimation at all $t$, i.e, a minimal reverse estimation, the reverse
estimation should be optimal. In the proof, a key point was that image of
m-affine curve is also m-affine. 

The integration (\ref{rld-divergence-int}) is computed by Hayashi:%
\begin{equation}
D^{R}(\rho ||\sigma )={\rm Tr}\rho \log \rho ^{\frac{1}{2}}\sigma
^{-1}\rho ^{\frac{1}{2}}.  \label{rld-divergence}
\end{equation}

\section{Monotone Divergence}

Let $D^{Q}(\rho ||\sigma )$ be a quantity which coincides with classical
divergence in the space of probability distributions, non-increasing by
application of a CPT map, and is additive,%
\[
D^{Q}(\rho _{1}\otimes \rho _{2}||\sigma _{1}\otimes \sigma _{2})=D^{Q}(\rho
_{1}||\sigma _{1})+D^{Q}(\rho _{2}||\sigma _{2}). 
\]
Then, in the almost the same way as monotone metric, we can conclude such
quantity is upper-bounded by $D^{R}(\rho ||\sigma )$, and 
lower-bounded by 
\[
D(\rho ||\sigma )
:=-{\rm Tr}\rho \left( \log \rho -\log \sigma \right) . 
\]
Assume that $D^{Q}(\rho ||\sigma )$ is monotone by a QC map, coincide with
the classical divergence for the probability distributions, and is
additive. It is known that there is a QC map such that the output
probability distributions $p_{\rho ^{\otimes n}}^{M}$ and $p_{\sigma
^{\otimes n}}^{M}$ satisfies, $D(\rho ||\sigma )=\frac{1}{n}D(p_{\rho
^{\otimes n}}^{M}||p_{\sigma ^{\otimes n}}^{M})+o\left( 1\right) $. This
implies 
\begin{eqnarray*}
D(\rho ||\sigma ) &=&\frac{1}{n}D^{Q}(p_{\rho ^{\otimes n}}^{M}||p_{\sigma
^{\otimes n}}^{M})+o\left( 1\right) \\
&\leq &\frac{1}{n}D^{Q}(\rho ^{\otimes n}||\sigma ^{\otimes n})+o\left(
1\right) \\
&=&D^{Q}(\rho ||\sigma )+o(1).
\end{eqnarray*}%
Here, tending $n\rightarrow \infty $, we have $D(\rho ||\sigma )\leq
D^{Q}(\rho ||\sigma )$ (This part is done by Hayashi).

On the other hand, assume that $D^{Q}(\rho ||\sigma )$ is non-increasing by
a CQ map, coincide with the classical divergence for the probability
distributions. Then, letting $(\Phi ,\{p_{\lambda }\})$ be an optimal
reverse estimation, 
\begin{eqnarray*}
D^{R}(\rho ||\sigma ) &=&D(p_{\rho }||p_{\sigma })=D^{Q}(p_{\rho
}||p_{\sigma }) \\
&\geq &D^{Q}(\rho ||\sigma ).
\end{eqnarray*}

\begin{theorem}
Assume that $D^{Q}(\rho ||\sigma )$ coincides with classical divergence for
the probability distributions. In addition, if $D^{Q}(\rho ||\sigma )$ is
additive and non-increasing by a QC map, 
\[
D(\rho ||\sigma )\leq D^{Q}(\rho ||\sigma ). 
\]%
On the other hand, if $D^{Q}(\rho ||\sigma )$ is non-increasing by a CQ map, 
\[
D^{Q}(\rho ||\sigma )\leq D^{R}(\rho ||\sigma ). 
\]
\end{theorem}

Can additivity assumption decrease the upper bound to the monotone
divergences ? This cannot be true, for $D^{R}(\rho ||\sigma )$ is additive.
On the other hand, if we remove the additivity assumption, the lower-bound
can be increased.

\section{Local reverse estimation of a multi-dimensional family}

A local reverse estimation can be recasted as follows. Under the constraint
of%
\[
A_{i}^{R}=U\widetilde{A_{i}^{R}}U^{\dagger },\quad \left[ \widetilde{%
A_{i}^{R}},\,\widetilde{A_{j}^{R}}\right] =0\quad (i,j=1,\cdots ,m), 
\]%
we minimize 
\begin{eqnarray}
&&\sum_{i,j}G_{ij}{\rm Tr}\sqrt{\rho }U\widetilde{A_{i}^{R}}\widetilde{%
A_{j}^{R}}U^{\dagger }\sqrt{\rho }  \nonumber \\
&=&\sum_{i,j}G_{ij}{\rm Tr}U^{\dagger }\rho U\widetilde{A_{i}^{R}}%
\widetilde{A_{j}^{R}},  \label{local-sim-mult-target}
\end{eqnarray}%
where $U$ is an isometry from ${\cal H}^{\prime }$ to ${\cal H}$, with 
$\dim {\cal H}^{\prime }\geq \dim {\cal H}$.

Here, note the analogy of this with the local state estimation, which gives
same result as the first order asymptotic theory. Assume we measure set of
observables $X^{i}$ to estimate $\theta ^{i}$, i.e., $X^{i}=\sum_{\kappa }%
\widehat{\theta _{\kappa }^{i}}\,M_{\kappa }$, where $\left\{ \widehat{%
\theta _{\kappa }^{i}};\kappa \right\} $ is an estimate of $\theta ^{i}$,
and $\left\{ M_{\kappa }\right\} $ is a POVM for a measurement used for the
estimation. Then, due to Naimark extension, we can find a set of observables 
$\widetilde{X^{i}}$ ($i=1,2,\cdots ,m$) with%
\[
X^{i}=U\widetilde{X^{i}}U^{\dagger },\quad \left[ \widetilde{X^{i}},\,%
\widetilde{X^{j}}\right] =0\quad (i,j=1,\cdots ,m), 
\]
and 
\[
{\rm Tr}U^{\dagger }\rho U\widetilde{X^{i}}\widetilde{X^{j}}=\sum_{\kappa
}\widehat{\theta _{\kappa }^{i}}\widehat{\theta _{\kappa }^{j}}{\rm Tr}%
\rho \,M_{\kappa }. 
\]%
Hence, establishing correspondence between $X^{i}$ and $A_{i}^{R}$, our
target function (\ref{local-sim-mult-target}) corresponds to the weighted sum
of the 'mean squared error' with the fixed set of observables. In other
words, the problem is reduced to optimization of measurement in quantum
estimation with the constraint $X^{i}=\sum_{\kappa }\widehat{\theta _{\kappa
}^{i}}\,M_{\kappa }$.

In particular, consider asymptotic exact reverse estimation with corrective
operation, i.e., the minimization of 
\[
\liminf_{n\rightarrow \infty }\frac{1}{n}\sum_{i,j}G_{ij}{\rm Tr}%
U^{\dagger }\rho ^{\otimes n}U\widetilde{A_{i}^{R,n}}\widetilde{A_{j}^{R,n}},
\]%
with the constraint 
\begin{eqnarray*}
\Phi \left( p_{\theta }^{n}\right)  &=&\rho _{\theta }^{\otimes n}, \\
\Phi \left( \partial _{i}p_{\theta }^{n}\right)  &=&\partial _{i}\rho
_{\theta }^{\otimes n}.
\end{eqnarray*}%
Define 
\begin{eqnarray*}
X_{i}^{n} &:&=\frac{1}{n}\left( \rho _{\theta }^{\otimes n}\right) ^{-\frac{1%
}{2}}\partial _{i}\rho _{\theta }^{\otimes n}\left( \rho _{\theta }^{\otimes
n}\right) ^{-\frac{1}{2}}. \\
&=&\frac{1}{n}\sum_{k=1}^{n}I\otimes \cdots \otimes A_{i}^{R}\otimes I\cdots
\otimes I.
\end{eqnarray*}%
Then, our target function is%
\[
\min_{U}\liminf_{n\rightarrow \infty }n\sum_{i,j}G_{ij}{\rm Tr}%
U^{\dagger }\rho ^{\otimes n}U\widetilde{X_{i}^{n}}\widetilde{X_{j}^{n}},
\]%
where $U$ runs over all isometry such that 
\[
X_{i}^{n}=U\widetilde{X_{i}^{n}}U^{\dagger },\quad \left[ \widetilde{%
X_{i}^{n}},\,\,\widetilde{X_{j}^{n}}\right] =0,\quad (i,j=1,\cdots ,\,m).
\]%
This corresponds to the asymptotic lower bound to the weighted sum of mean
square error of corrective measurements. Hence, the minimum is given using
so-called Holevo bound. For we have 
\[
{\rm Tr}\rho A_{i}^{R}A_{j}^{R}=J_{i,j}^{R},
\]%
due to Holevo bound, we have 
\begin{eqnarray}
&&\min_{U}\liminf_{n\rightarrow \infty }
n\sum_{i,j}G_{ij}{\rm Tr}U^{\dagger }
\rho ^{\otimes n}U\widetilde{X_{i}^{n}}\widetilde{X_{j}^{n}} \nonumber\\
&=&{\rm Sp}G\Re J_{\theta }^{R}+{\rm Spabs}G\Im J_{\theta }^{R}, 
\label{l-rest-multi}\\
&=&\min \left\{ {\rm Sp}GJ\,;\,J\geq J_{\theta }^{R}\right\} .
\nonumber
\end{eqnarray}%
Note%
\begin{eqnarray*}
\Im J_{\theta }^{R} 
=-\frac{1}{2}{\rm Tr}\rho _{\theta }\left[ L_{i}^{R},\,L_{j}^{R}\right]
,
\end{eqnarray*}%
and this quantity is a measure of non-commutativity of RLD's. If this
quantity is larger, we need more classical Fisher information than the real
part of RLD. 

On the other hand, if the given state family is $D$-invariant
in Holevo's sense, the bound corresponding to the estimation is given, 
\begin{eqnarray*}
& &\max \left\{ {\rm Sp}GJ\,;\,J\leq J_{\theta }^{R}\right\}  \\
&=&{\rm Sp}G\Re J_{\theta }^{R}-{\rm Spabs}G\Im J_{\theta }^{R},
\end{eqnarray*}%
and the bound is achievable.
This is smaller than the reverse estimation bound by 
${\rm Spabs}G\Im J_{\theta }^{R}$.

\begin{example} {\bf (Gaussian state family)}
A Gaussian state family is defined by%
\[
\rho _{\theta }=\int \frac{{\rm d}p{\rm d}q}{2\pi \sigma _{p}\sigma
_{q}}e^{-\frac{1}{2\sigma ^{2}}\left\{ (q-\theta ^{1})^{2}+(p-\theta
^{2})^{2}\right\} }\left\vert p,q\right\rangle \left\langle p,q\right\vert 
\]
This definition itsefl gives a global reverse estimation
such that the coherent state $\left\vert p,q\right\rangle$
is according to the Gaussian distribution with the variance $\sigma^2$
and the mean $\theta=(\theta^1,\theta^2)$.
Its input Fisher information is $J=\sigma ^{-2}{\rm I}$,
and 
\[
 {\rm Sp}J=2\sigma ^{-2}.
\]
This in fact is optimal:
\[
{\rm Sp}\Re J_{\theta }^{R}+{\rm Spabs}\Im J_{\theta }^{R}
=2\sigma ^{-2},
\]
where
\[
J^{R}=\frac{1}{\left( \sigma ^{2}+\hbar \right) \sigma ^{2}}\left[ 
\begin{array}{cc}
\sigma ^{2}+\frac{\hbar }{2} & -i\hbar /2 \\ 
i\hbar /2 & \sigma ^{2}+\frac{\hbar }{2}%
\end{array}%
\right] .
\]

\end{example}


\begin{thebibliography}{9}
{\small 
\bibitem{Fujiwara} A.~Fujiwara, QCMC '96 (1996)

\bibitem{Hayashi:2005:book} M.~Hayashi, Quantum Information theory, Springer
Ver-lag, to appear

\bibitem{Nagaoka} S.~Amari, H.~Nagaoka,
Methods of Information Geometry,
American Mathematical Society@(2001)

\bibitem{Nagaoka2} 
H. Nagaoka, gOn the Parameter Estimation Problem for Quantum Statistical Mod-
els,h SITAf89, 577-582 Dec. (1989)

\bibitem{Petz} D. Pets, 
Monotone metrics on matrix spaces, Linear Algebra Appl. 244
(1996), 81.96.}
\end{thebibliography}
\end{document}